\documentstyle[11pt,epsfig]{article}
\begin{document}

\begin{center}
{\bf STRICTLY ISOSPECTRAL POTENTIALS FROM EXCITED QUANTUM STATES}
\end{center}

\bigskip

\begin{center}
R. KLIPPERT$^{1,2}$ and H.C. ROSU$^{1,3}$ 


{\scriptsize $^1$ International Center for Relativistic Astrophysics, 
Piazzale della Repubblica 10,\\[-1.2ex] 65100 Pescara, Italy}

{\scriptsize $^2$ Brazilian Center for Research in Physics, R. Dr.
Xavier Sigaud 150 Urca, 22290-180 \\[-1.2ex]Rio de Janeiro RJ, Brazil}\\[-0.7ex]

{\scriptsize $^3$ Instituto de F\'{\i}sica, Univ.\ de Guanajuato, 
Apdo Postal E-143, Le\'on, Gto, Mexico}

\end{center}

\bigskip
\bigskip

\noindent
{\bf Abstract.} - The strictly nonrelativistic isospectral scheme 
based on the general Riccati solution and Darboux transformation function
corresponding to excited states is presented on the following examples: the harmonic 
oscillator, the square well, and the hydrogen atom.  

\bigskip 
\bigskip 

\bigskip 
\bigskip 
\bigskip

Supersymmetric quantum mechanics (SUSY QM) \cite{HR} is based on Riccati equations of the form
\begin{equation} \label{e1}
\frac{dy}{dz}+y^2=\tilde{V_{1}}(z)+\epsilon ~,
\end{equation}
where $\tilde{V_{1}}(z)$ is the initial 
Schr\"odinger potential and $\epsilon < 0$ is the so-called factorization 
energy that for convenience we shall absorb in the potential, i.e., we shall redefine the 
initial potential as $V_1=\tilde{V_1}+\epsilon$. The SUSY partner Riccati equation will be 
\begin{equation} \label{e2}
-\frac{dy}{dz}+y^2=V_{2}(z)~,
\end{equation}
where the partner potential $V_{2}(z)$ is Darboux `isospectral' with 
respect to $V_{1}(z)$, i.e.
\begin{equation} \label{e3}
V_{2}(z)=V_{1}(z)-2D^2[\ln (\psi(z))]~,
\end{equation} 
where $D=\frac{d}{dz}$ and $\psi$ is a particular solution of the Schr\"o\-din\-ger equation $D^2\psi-V_1\psi=0$. The only difference between the spectra of $V_{2}$ and $V_{1}$ is that 
the energy level corresponding to the wavefunction $\psi (z)$ used as transformation function
is missing from the spectrum of $V_{2}$.

In 1984, Mielnik introduced a SUSY QM scheme based on the general 
Riccati solution with application to the harmonic oscillator 
case \cite{M}. In the latter approach, 
one gets a one-parameter family of Darboux strictly isospectral
potentials with respect to $V_{1}$ given by
\begin{equation} \label{e4}
V_{1}(z;\lambda)=V_{1}(z)-2D^2[\ln (I(z)+\lambda)]~,
\end{equation}
where $I(z)=\int _{l}^{z}\psi^2(x)dx$. Moreover,
\begin{equation} \label{e5}
\psi(z,\lambda)=\frac{\sqrt{\lambda (\lambda +1)}\psi(z)}{\int _{l}^{z}\psi^2(x)dx+\lambda}
\end{equation}
is the modulated Schr\"odinger mode implied by this scheme in which 
the Riccati integration constant $\lambda$ is kept as a free parameter. 
The inferior limit $l$ of the integral $I(z)$ is zero for the radial problems, 
$-\infty$ for the full-line problems, and $-L/2$ for the square well case (see below).
The factor $\sqrt{\lambda (\lambda +1)}$ is a normalization constant of the strictly
isospectral SUSY modes that interestingly does not depend on the quantum numbers.
In general,
to get continuous solutions $\psi(z,\lambda)$, one should avoid 
a zero denominator. This leads to conditions on the possible values 
of $\lambda$.
On the other hand, if one works with polynomial solutions, which is the 
usual case for the discrete spectrum of an exactly solvable quantum
problem, there will be 
singularities in the logarithmic derivative of $\psi$. SUSY partner potentials based 
on the $n^\prime$th excited state of a Schr\"odinger discrete spectrum problem
split in $n+1$ branches separated by the $n$ singularities of the logarithmic 
derivative. Such problems have been considered by Robnik \cite{Rob}.

In the following, we focus on Robnik's results from the perspective of the 
strictly isospectral scheme, namely instead of working with Eq.~(\ref{e3}) as Robnik did, we 
make use of Eqs.~(\ref{e4}) and (\ref{e5}). This is also equivalent to saying that we work with factorization
energies $\epsilon =\epsilon _{n}$ in Eq.~(\ref{e1}) and $\psi =\psi _{n}$ as transformation function in Eq.~(\ref{e4}),
where $\epsilon _{n}$ is an eigenvalue of the discrete spectrum of the exactly
solvable problem and $\psi _{n}$ is the corresponding eigenfunction. Our applications are
the following.

\bigskip

\noindent
(i) \underline{{\rm The one-dimensional harmonic oscillator}}

\bigskip

The initial potential is 
\begin{equation} \label{e6b}
V_{1}(z)=
z^2-(2n+1)~,
\end{equation}
for $n\in 0,1,2,...$. 
The wave functions are given in terms of the Hermite polynomials in the form:
\begin{equation} \label{e6}
\psi _n =\frac{1}{\sqrt{2^n n! \sqrt{\pi}}} H_{n}(z)\exp (-z^2/2)
\end{equation}

\begin{figure}[htbp]
\leavevmode
\centerline{
\centering
\epsfxsize=65ex
\epsfbox{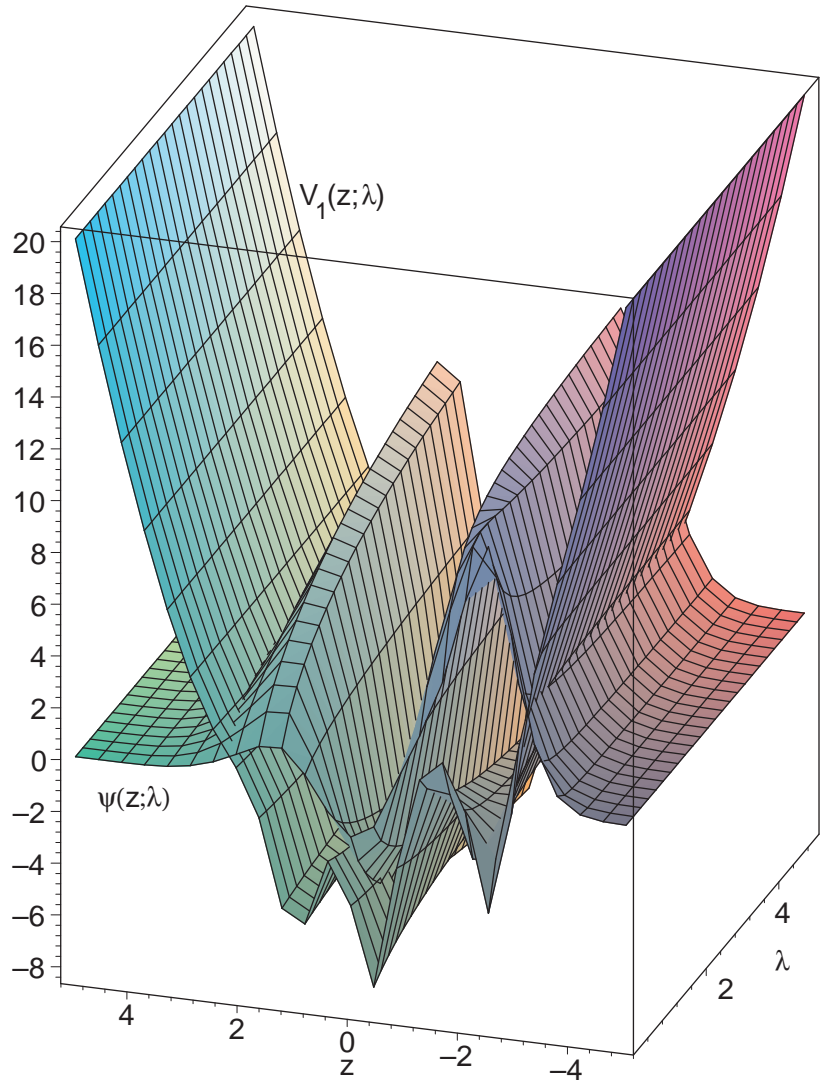}}
\caption{The strictly isospectral potential and the deformed wavefunction
(multiplied by 10 for the sake of visualization) 
obtained using the $n=2$ excited state of the harmonic oscillator as 
Darboux transformation function.}
\label{fig1}
\end{figure}

\begin{figure}[htbp]
\leavevmode
\centerline{
\centering
\epsfxsize=65ex
\epsfbox{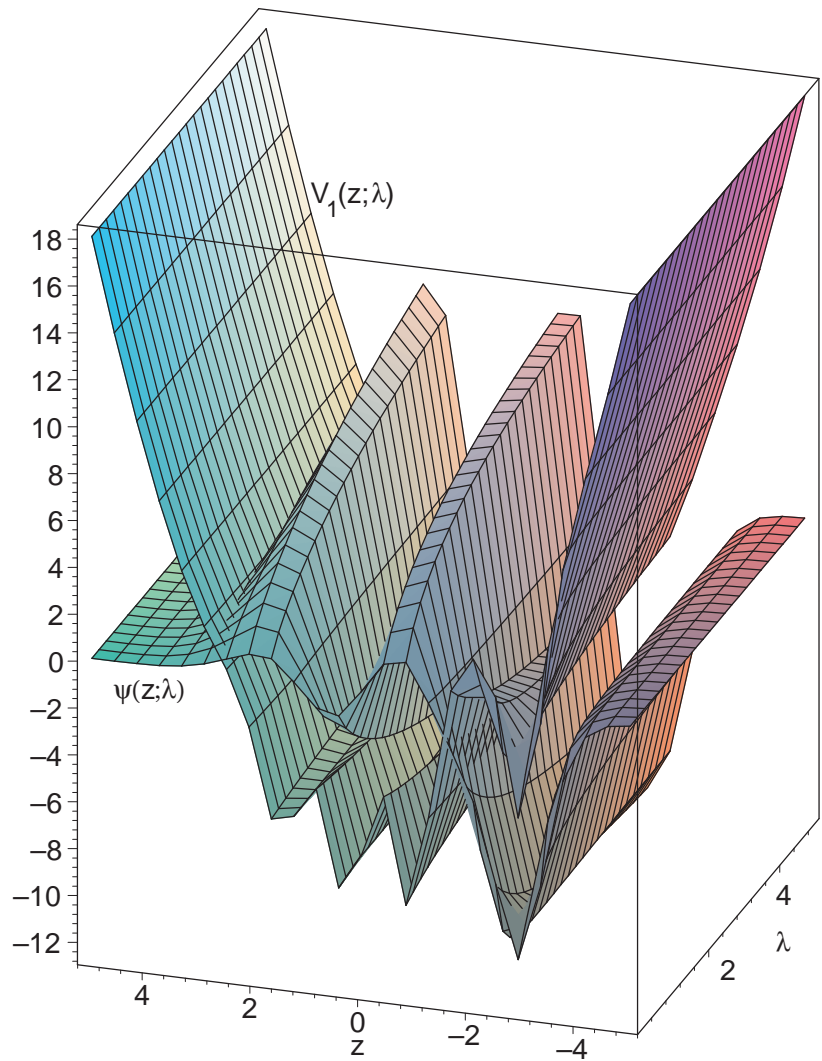}}
\caption{The strictly isospectral potential and the deformed wavefunction
(with the same scale factor of 10) 
obtained using the $n=3$ excited state of the harmonic oscillator as 
Darboux transformation function.}
\label{fig2}
\end{figure}

We present plots of Eqs.~(\ref{e4}) and (\ref{e5}) corresponding to $n=2$ and $n=3$ in
Figs.~\ref{fig1} and \ref{fig2}, respectively.
The examination of the plots shows that there are 
no singularities in the one parameter family of potentials for the 
allowed range of the $\lambda$ parameter. Instead, there are $n+1$
wells, where $n$ is the number of nodes of the excited state on 
which the strictly isospectral SUSY construction is based. Thus, we have at our disposal a definite method
of producing multiple-well oscillator potentials of exactly solvable type.

\bigskip





\bigskip

\noindent
(ii) \underline{{\rm One-dimensional square well of length $L$}}

\bigskip

The initial potential is:
\begin{equation} \label{e8b}
V_1(z)=\left\{
\begin{array}{l@{\hspace{2em}}l}
-\frac{\pi ^2}{L^2}n^2 \quad & {\rm for} -L/2 \leq z \leq L/2~, \\ 
V_1(z)=+\infty & {\rm otherwise}~,
\end{array}
\right.
\end{equation}
where $n$ belongs to the set $n=1,2,...$.
The wave functions within the well are:
\begin{equation} \label{e8}
\psi _n =\sqrt{\frac{2}{L}} \cos (\frac{n\pi z}{L})\quad {\rm for\ odd} \quad n~,
\end{equation}
 and:
\begin{equation} \label{e8c}
\psi _n =\sqrt{\frac{2}{L}} \sin (\frac{n\pi z}{L})~\quad {\rm for\ even} \quad n~.
\end{equation}

\begin{figure}[htbp]
\leavevmode
\centerline{
\centering
\epsfxsize=65ex
\epsfbox{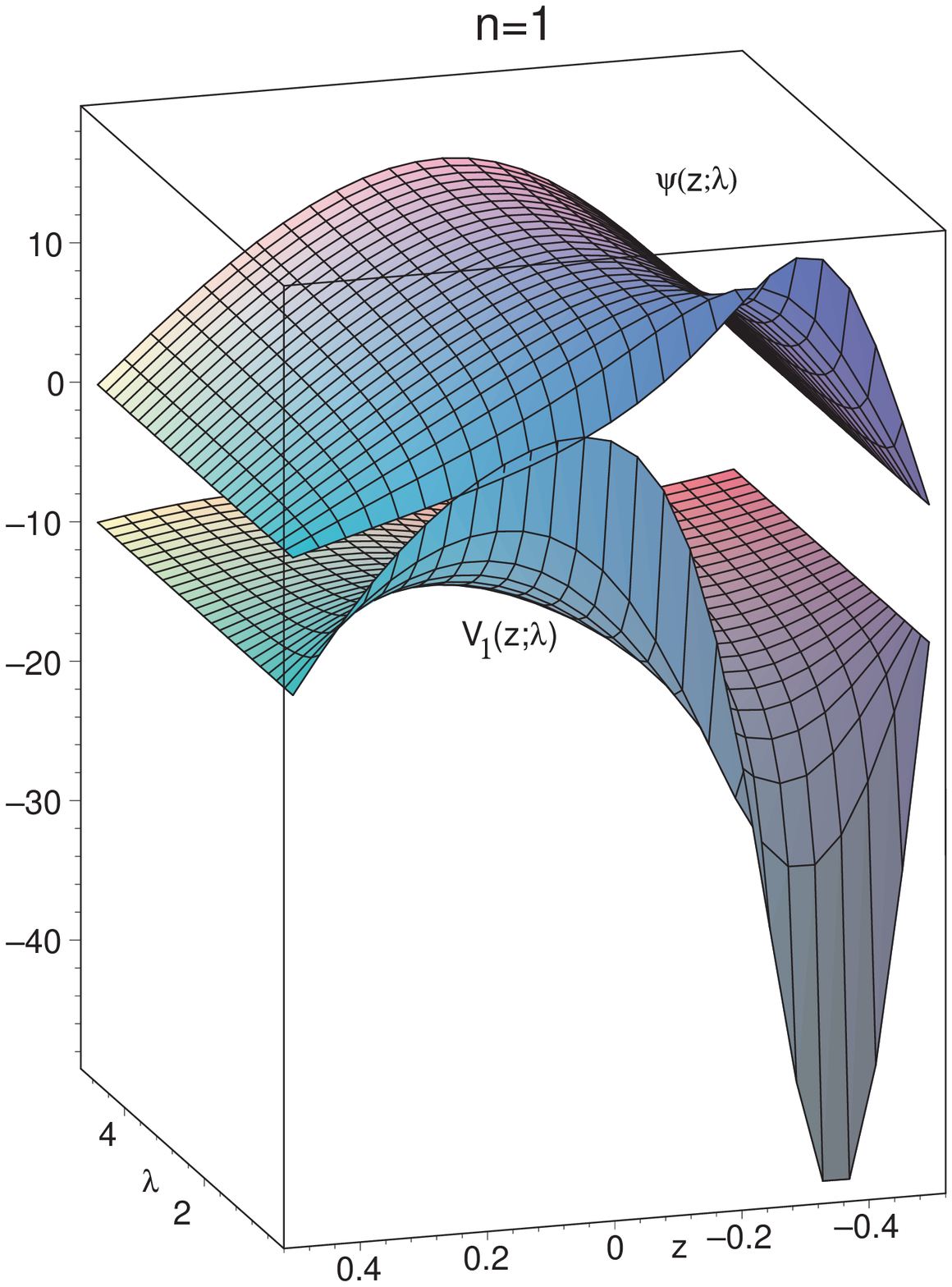}}
\caption{The strictly isospectral potential and the deformed wavefunction
$\times$ 10  
obtained using the $n=1$ excited state of the square well as 
Darboux transformation function.}
\label{fig3}
\end{figure}

\begin{figure}[htbp]
\leavevmode
\centerline{
\centering
\epsfxsize=65ex
\epsfbox{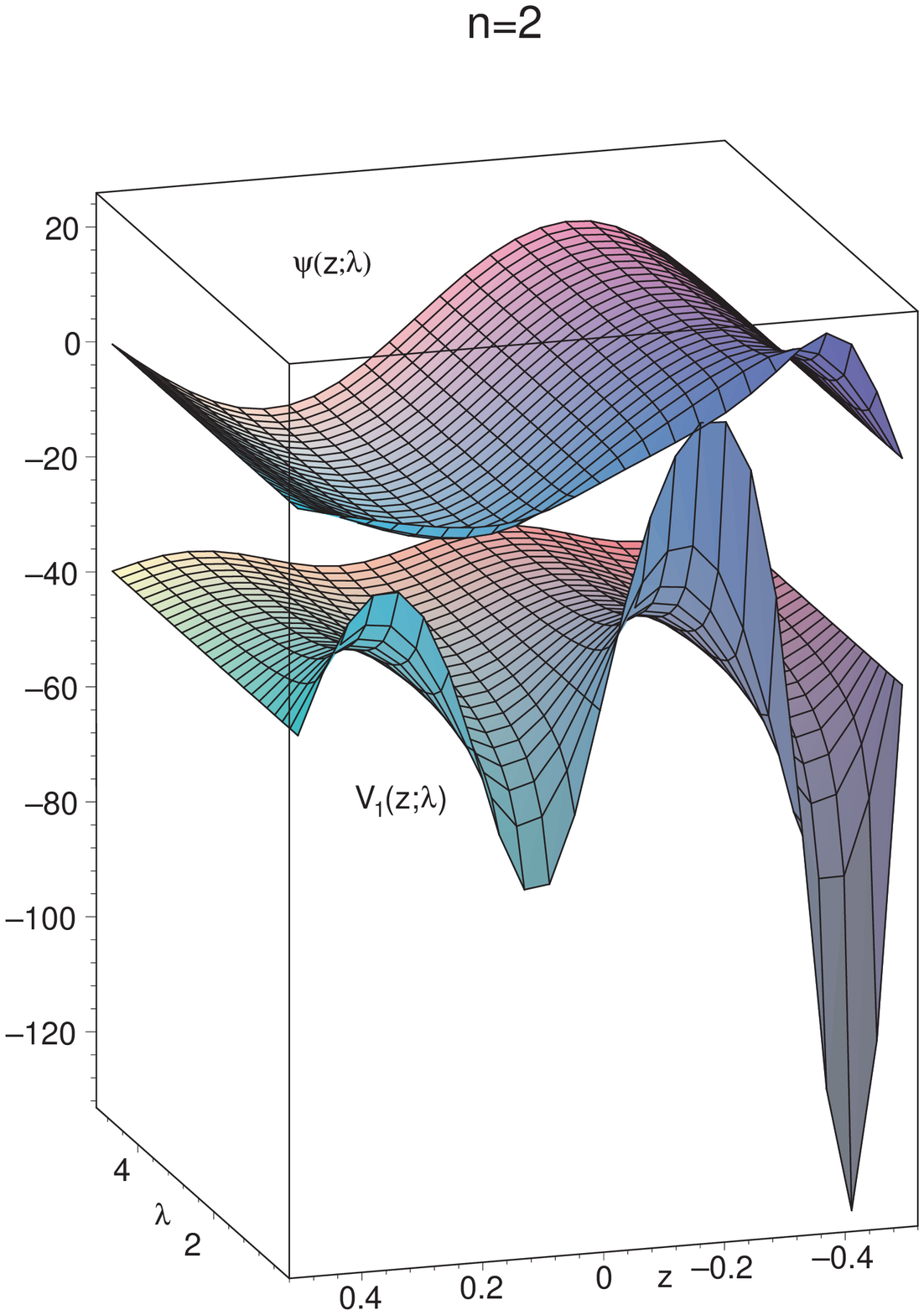}}
\caption{The strictly isospectral potential and the deformed wavefunction $\times\,10$
obtained using the $n=2$ excited state of the square well as 
Darboux transformation function.}
\label{fig4}
\end{figure}

Plots of the strictly isospectral formulas for this case are given in Figs.~\ref{fig3} and \ref{fig4} for an odd case and an even case, respectively.

\bigskip

\bigskip

\noindent
(iii) \underline{{\rm The three-dimensional Kepler problem}}

\bigskip
The initial potential is
\begin{equation} \label{e9b}
V_{1}(z)=-\frac{2}{z}+\frac{l(l+1)}{z^2}-\frac{4}{(n_r+l+1)^2}~,
\end{equation}
for $n_r$ from the set $n_r=0,1,2,...$, and where the centrifugal barrier has been also
included.

\noindent
The radial wave functions to be used as Darboux transformation functions are
\begin{equation} \label{e9}
R _{nl} =-\frac{2}{n^2}\sqrt{\frac{(n-l-1)!}{2n[(n+l)!]^3}}z^l e^{- z/2}L_{n+l}^{2l+1}
(z)
\end{equation}
where $z=2r/n$ and the Bohr radius has been chosen as unit.

\begin{figure}[htbp]
\leavevmode
\centerline{
\centering
\epsfxsize=65ex
\epsfbox{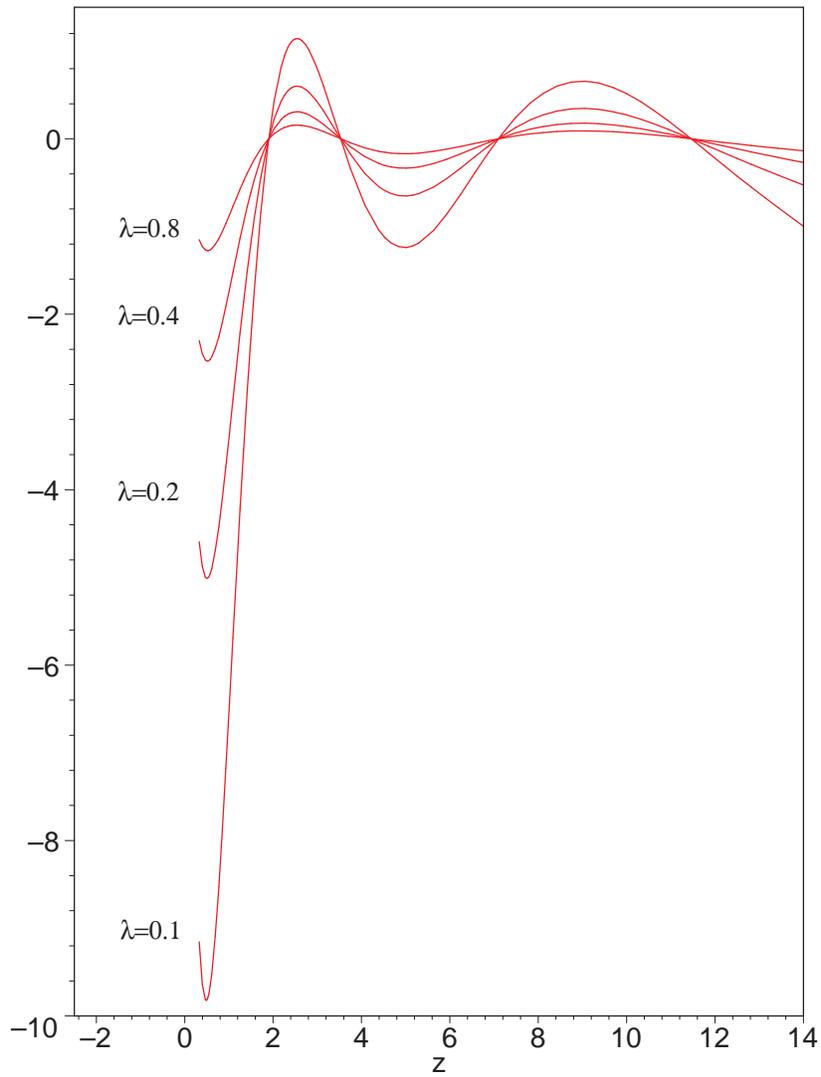}}
\caption{The percentual difference $100(V_{1}(z; \lambda)-V_{1}(z))/V_{1}(z)$ between the 
strictly isospectral Coulomb potential and the Coulomb potential when the hydrogen radial
state $R_{30}$ is used as Darboux transformation function for four 
values of the parameter $\lambda$.}
\label{fig5}
\end{figure}

\begin{figure}[htbp]
\leavevmode
\centerline{
\centering
\epsfxsize=65ex
\epsfbox{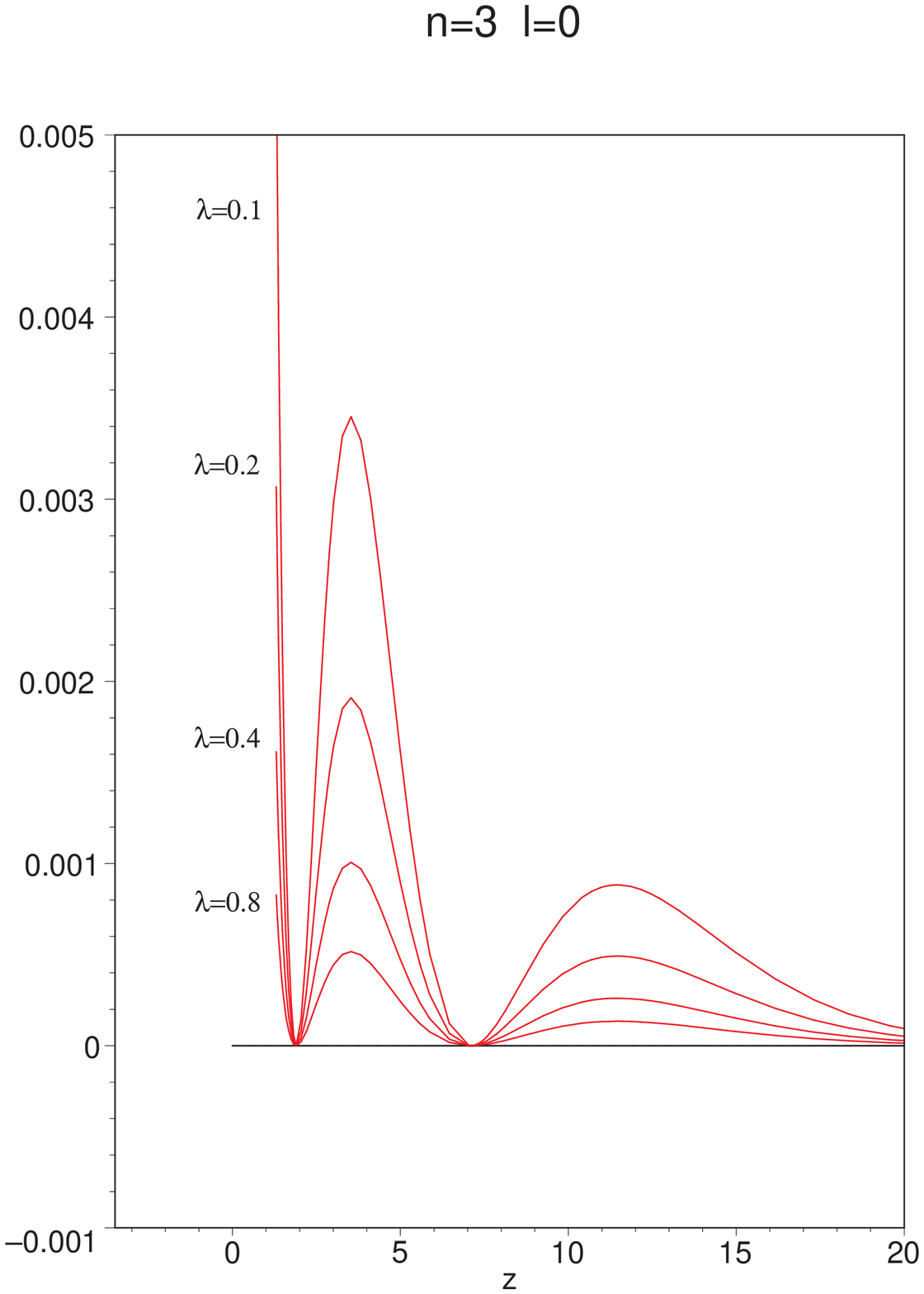}}
\caption{The difference between the probability densities $\psi ^{2}(z;\lambda)-
\psi ^{2}(z)$ for the same case as in Fig.~5.}
\label{fig6}
\end{figure}

\begin{figure}[htbp]
\leavevmode
\centerline{
\centering
\epsfxsize=65ex
\epsfbox{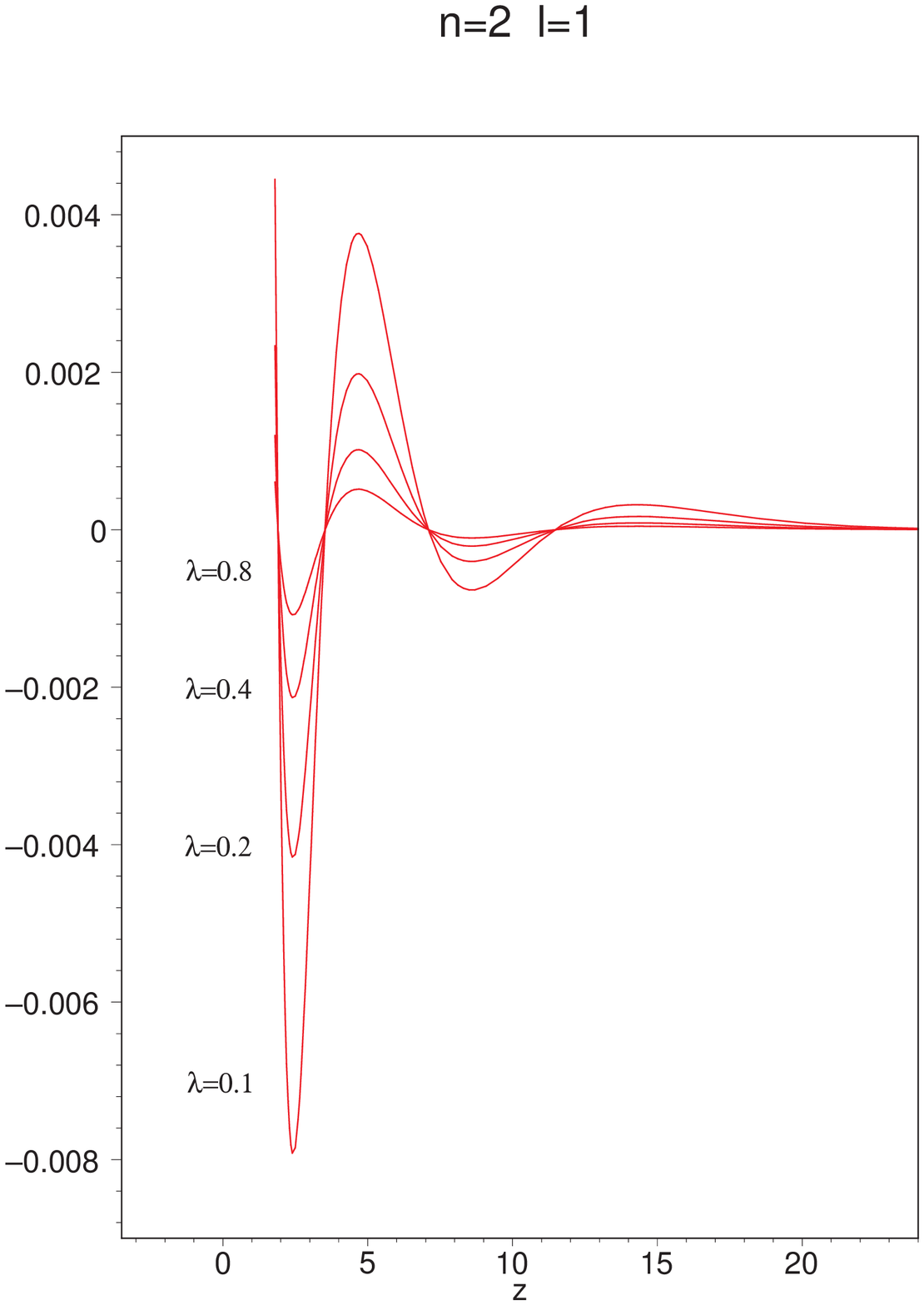}}
\caption{The difference between the strictly isospectral 
Coulomb potential and the Coulomb potential $V_{1}(z; \lambda)-V_{1}(z)$
when the hydrogen radial
state $R_{21}$ is used as Darboux transformation function for the same four 
values of the parameter $\lambda$.}
\label{fig7}
\end{figure}

\begin{figure}[htbp]
\leavevmode
\centerline{
\centering
\epsfxsize=65ex
\epsfbox{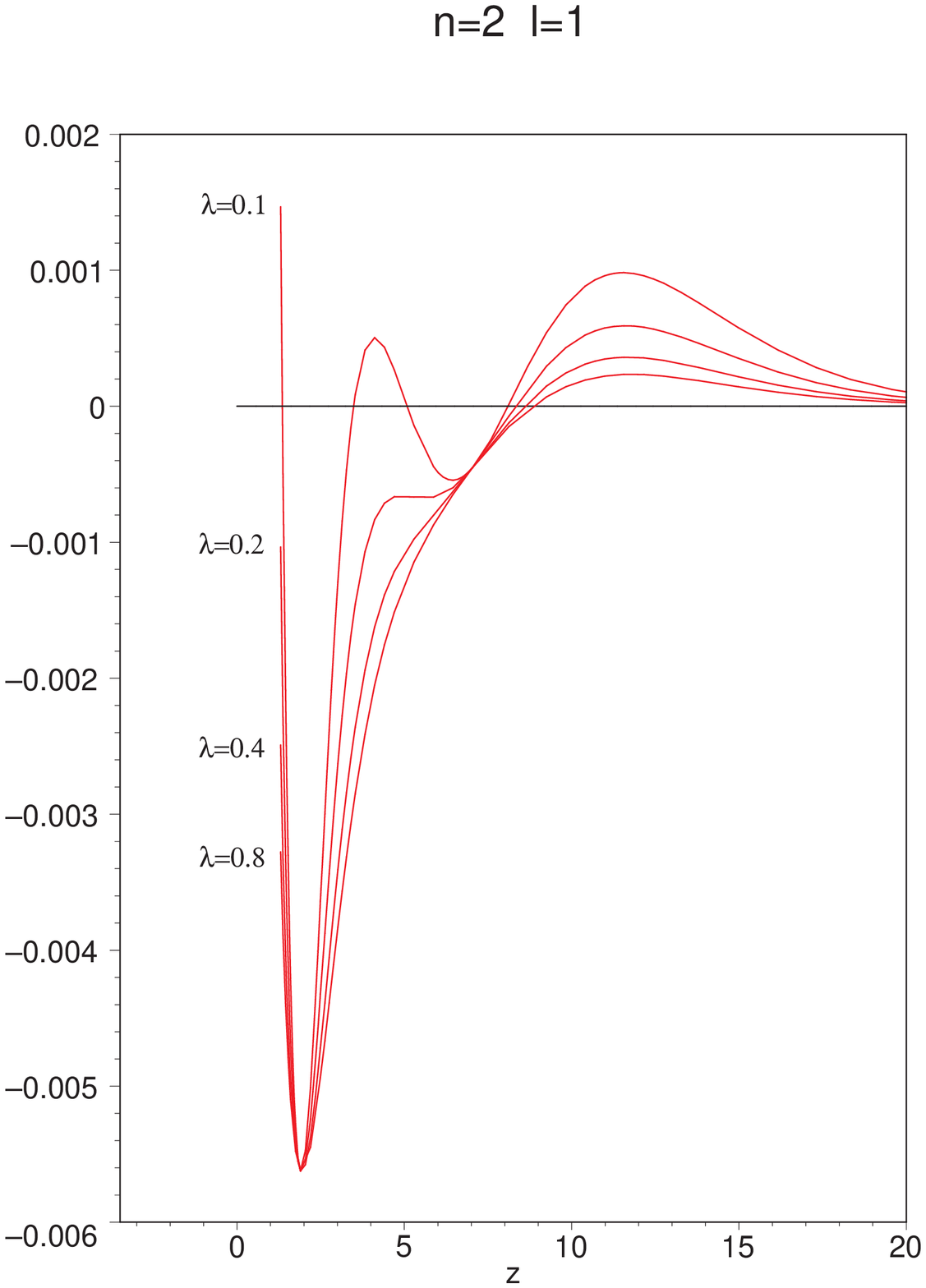}}
\caption{The difference between the probability densities $\psi ^{2}(z;\lambda)-
\psi ^{2}(z)$ for the same case as in Fig.~7.}
\label{fig8}
\end{figure}

In Figs.~\ref{fig5}--\ref{fig6} and \ref{fig7}--\ref{fig8} we present the result of the strictly isospectral 
construction based on the hydrogen radial wavefunctions $R_{30}$ and $R_{21}$, respectively. 
For the 
employed range of the $\lambda$ parameter, there is a small strictly isospectral effect
extending in the Rydberg region till about 20 Bohr radii. On the other hand, although 
not shown in the plots, we can report that in the $z=0$ nuclear region the efect can be as 
big as 25\%.

We recall that for this case a similar approach has been used by Fern\'andez \cite{Fer}. However, his procedure cannot be applied to the 
$l=0$ cases, whereas the way we use the strictly isospectral Darboux scheme covers the full hydrogen spectrum.

Finally, we notice that it is still an open issue to what definite physical situation the 
strictly isospectral scheme does apply. There are hints 
in the literature \cite{hints} indicating that $\lambda$
can be used as a measure of confining effects on the 
quantum spectra.

\clearpage
\section*{Acknowledgements}

We would like to thank Prof.\ R. Ruffini for hospitality at the ICRA center in Pescara.
RK aknowledges Brazilian CAPES foundation for a grant.



\end{document}